\documentclass[%
footinbib,
aip,
 amsmath,amssymb,
 reprint,%
]{revtex4-2}

\usepackage[utf8]{inputenc}
\usepackage[T1]{fontenc}
\usepackage{mathptmx}
\usepackage{etoolbox}

\usepackage{color}
\usepackage{graphicx}
\usepackage[normalem]{ulem}
\usepackage{soul}
\usepackage{dcolumn}
\usepackage{bm}
\definecolor{darkgreen}{RGB}{20,150,40}

\usepackage{comment}
\usepackage{amsfonts}
\usepackage{amsmath}
\usepackage{stackengine}
\usepackage{wrapfig}
\usepackage[x11names]{xcolor}
\usepackage{vwcol}
\usepackage{appendix}

\makeatletter
\def\@email#1#2{
 \endgroup
 \patchcmd{\titleblock@produce}
  {\frontmatter@RRAPformat}
  {\frontmatter@RRAPformat{\produce@RRAP{*#1\href{mailto:#2}{#2}}}\frontmatter@RRAPformat}
  {}{}
}
\makeatother

\begin{document}

\preprint{AIP/123-QED}

\title{Nonlinear interactions of ion acoustic waves explored using fast imaging decompositions}

\author{Simon Vincent$^{1,2}$}
\author{Vincent Dolique$^{1}$}
\author{Nicolas Plihon$^{1}$}
\affiliation{$^{1}$ Univ Lyon, ENS de Lyon, CNRS, Laboratoire de Physique, F-69342 Lyon, France \\
$^{2}$ École Polytechnique Fédérale de Lausanne (EPFL), Swiss Plasma Center (SPC), CH-1015 Lausanne, Switzerland}

\date{\today}
             
\begin{abstract}
Fast camera imaging is used to study ion acoustic waves propagating azimuthally in a magnetized plasma column. The high speed image sequences are analyzed using Proper Orthogonal Decomposition and 2D Fourier Transform, allowing to evaluate the assets and differences of both decomposition techniques. The spatio-temporal features of the waves are extracted from the high speed images, and highlight energy exchanges between modes. Growth rates of the modes are extracted from the reconstructed temporal evolution of the modes, revealing the influence of ion-neutral collisions as  pressure increases.  Finally, the nonlinear interactions between modes are extracted using bicoherence computations, and show the importance of interactions between modes with azimuthal wave numbers $m$, $m-1$ and $-1$, with $m$ an integer. 
\end{abstract}


\maketitle

\section{Introduction}

The propagation of ion sound waves or ion acoustic waves is ubiquitous in plasmas and their non-linear interactions, possibly leading to ion acoustic turbulence, is a widespread energizing process in plasma physics. The nonlinear evolution of ion acoustic waves (IAW) generically leads to instabilities and the development of non-linear structures. For instance, IAW have long been observed in the solar wind, and related to the anisotropy of the 
electron distribution function~\cite{Gurnett_1978}. In this context, IAW are driven unstable when the ratio of the electron to ion temperature is larger than unity, as observed by the Helios spacecraft~\cite{Gurnett_1979}, and very recently for oblique IAW by Parker Solar Probe~\cite{Mozer_2021}. Heating of energetic particles from ion acoustic turbulence was also proposed in the context of polar aurorae~\cite{Wahlund_1994}.
The non-linear evolution of ion acoustic waves into strongly non-linear structures such as solitons~\cite{Ikezi_1973} or double layers has been reported in electro-positive plasmas~\cite{Sato_1980} or electronegative plasmas~\cite{Plihon_2011}, for which two branches of ion acoustic waves exist~\cite{DAngelo_1966_2}. In the context of bounded plasmas, IAW excited in sheaths may affect particle transport at low pressure~\cite{Baalrud_2016} or 
lead to strong ion heating~\cite{Beving_2021} when the ratio of the electron to ion temperature is larger than unity. IAW may also be useful tools to probe sheath criteria in multiple ion plasmas~\cite{Lee_2006,Oksuz_2008,Hala_2001}. 
Technological plasmas may also trigger IAW, that, in return, affect their operation, as reported for Hollow Cathodes~\cite{Jorns_2017,Tsikata_2021}, Hall thrusters~\cite{Katz_2016} and diverging magnetic nozzle thrusters~\cite{Doyle_2020}.

In this article, we report on the observation of localized ion acoustic waves in a magnetized plasma column using high speed camera imaging. Our observations thus shed new light on the ion acoustic activity that has been previously reported in similar configurations~\cite{Boswell_1976, Virko_2003, Corr_2004,  Belov_2006, Lorenz_2005, Kramer_2007, Corr_2009}. We do not investigate the origin of the IAW from parametric instability or waves interactions here, as was done in these previous investigations, but we analyse the spatio-temporal characteristics of the IAW using mode decomposition from high-speed imaging. The IAW nonlinear interactions are quantitatively highlighted by means of bicoherence computations. 

The article is organized as follows. The experimental set-up is introduced in Sec.~\ref{section_set-up}, the analysis of fast camera measurements by mode decomposition techniques is presented in Sec.~\ref{section_image_analysis}. In particular, we highlight the differences and complementarities of two different mode decompositions, namely Proper Orthogonal Decomposition and 2D Fourier Transform.  In section~\ref{section_identification} the waves observed by camera imaging are identified to be IAW from the waves phase velocities. Finally the non-linear modes interactions are characterized and their nonlinear aspect is exhibited in section \ref{section_NL_interactions} and conclusions are drawn in section~\ref{sec:ccl}.

\section{Experimental set-up and diagnostics}
\label{section_set-up}

\subsection{Experimental set-up}
\label{subsection_set-up}

\begin{figure}
    \includegraphics[width = 0.98\columnwidth, trim={0in 0in 0in 0in},clip]{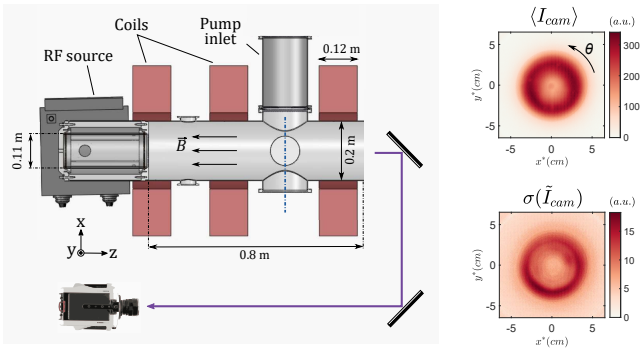}
    \caption{Left: sketch of the experimental set-up. The dashed blue line indicates where probe profiles are measured.  Right: mean image of the light intensity collected by the camera (up), and standard deviation of the fluctuations around this average value (bottom), for $p_0=1$~mTorr.}
    \label{VKP_picture}
\end{figure}

The experimental set-up~\cite{Plihon_2015} consists in a $20$~cm diameter, $1$~m long stainless steel cylindrical chamber containing an argon plasma generated by a $1$~kW, 13.56 MHz radio frequency source. The coordinates will be referenced using a cartesian coordinate system, where $z$ denotes the axial direction (see Fig.~\ref{VKP_picture}). A cylindrical coordinate system $(r,\theta,z)$ will be used for the reconstruction of rotating modes. The plasma base pressure $p_0$ in the chamber is regulated at a fixed value, between $0.8$~mTorr and $2$~mTorr by steps of $0.1$~mTorr. The plasma is created by an inductive source around a $11$~cm diameter borosilicate tube connected at one end of the chamber (at $z=0$~cm). Three coils placed along the steel cylinder generate an axial magnetic field that confines the plasma (see Fig.~\ref{VKP_picture}). This magnetic field is not perfectly homogeneous along $z$. For a current in the coils set here at $100$~A, it has an averaged amplitude along the $z$-axis of $B=170$~G. 

Radial profiles of the plasma density $n$, the electron temperature $T_e$ and the plasma potential $\Phi_p$ are performed using a 5-tips probe~\cite{Tsui_1992, Vincent_2022} and an emissive probe respectively. The probes were inserted radially at  $z=49$~cm (see blue dashed line in Fig.~\ref{VKP_picture}). To keep the whole apparatus in a steady thermal state, the operation of the plasma is pulsed: the plasma is sustained over typically 5 seconds, during which data are acquired, with a repetition period of typically 30~s. The experiment is fully automated to allow high repeatability and reproducibility of the plasma. The level of shot to shot reproducibility was $\pm 0.6\%$ for the ion saturation current of a Langmuir probe, with a standard deviation of $0.2\%$ (estimated from a series of 40 shots at the plasma column center). Radial scans of the plasma parameters were performed sequentially: each spatial point has been acquired during one plasma-pulse, and the probe is translated between two pulses, from the center of the plasma column ($r=0$~cm) to its edge ($r=10$~cm).

The results presented in this article mainly rely on high-speed imaging of the plasma emitted light, performed through a DN 200 borosilicate window closing the chamber at $z=80$~cm (opposite to the source tube). A Phantom v2511 camera is placed along the $z$-axis, $3.5$~m away from this window, and the light intensity naturally radiated by the plasma $I_{cam}$ is captured at $200$~kfps with a resolution of $256 \times 256$~px. A filter around $750 \pm 5$~nm is used in order to restrict the collected light to a single ArI spectral line. Examples of the mean intensity $\langle I_{cam}\rangle$ and fluctuation standard deviation $\sigma(\tilde{I}_{cam})$ images are shown in Fig.~\ref{VKP_picture}. Note that the plasma column is not perfectly axisymmetric, and the fluctuations are of the order of $10\%$ of the mean amplitude.
Note also that the depth of field of the optical set-up being of the order of the length of the chamber (with a camera objective aperture set at f/4, and a focal length of $135$~mm), the light intensity recorded by the camera is actually the result of an integration along the $z$-axis.
Due to the magnetic field ripple and to the parallax, the direct comparison of the probes' measurements (at $z=49$~cm) and the camera images (where light is integrated along $z$), is not relevant. We thus introduce a distorted space ($x^*$, $y^*$, $z$) in which the camera lines of sight are parallel (see Ref.~\cite{Vincent_2022} for more details). The camera images are hence observed in the plane ($x^*$, $y^*$), that may also be referenced as $(r^*,\theta)$ in a polar coordinate system.

\subsection{Radial profiles of the plasma parameters}
\label{subsection_mean-profiles}

\begin{figure}
    \centering
    \includegraphics[width = .98\columnwidth, trim={0in 0in 0in 0in},clip]{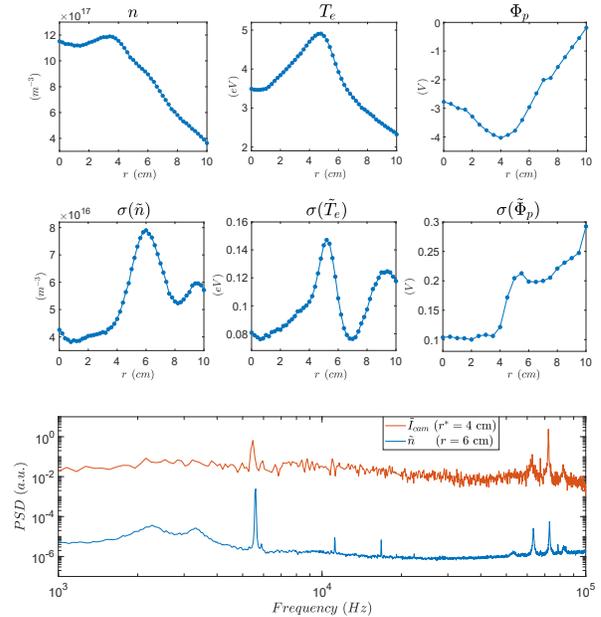}
    \caption{Radial profiles of the density, electron temperature, and plasma potential mean values (top row) and fluctuations (middle row) at $p_0 = 1$~mTorr. Plasma parameters measurements were made with 5-tips ($n$, $T_e$) and emissive ($\Phi_p$) probes. Bottom: Power Spectral density of the plasma density and light intensity fluctuations (computed from the average value of a $5 \times 5$ pixel box on the images).}
    \label{n_Te_Vp_probe}
\end{figure}

The top row of Fig.~\ref{n_Te_Vp_probe} displays the radial profiles of the plasma density, electron temperature and plasma potential as a function of $r$ for a pressure $p_0=1$~mTorr. As previously introduced, these profiles cannot be directly compared to the images in the $(r^*,\theta)$ plane. However, assuming axisymmetry and invariance of the plasma parameter along magnetic field lines, a synthetic integration process detailed in Ref~\cite{Vincent_2022} allows to map the probe measurements along r (at $z=49$~cm) to the images expressed along $r^*$, enabling quantitative comparison. The density is approximately constant in a core region of the plasma for $r \leq 4$~cm  ($r^* \leq 3$~cm) and then decreases towards the edge. A clear peak can be seen around $r=4.5$~cm ($r^* \sim 3$~cm) for the electron temperature, produced by the higher ionization rate of the RF inductive source close to the wall at $r=5.5$~cm, $z \sim -10$~cm. This higher temperature is also responsible for the higher light emission observed on the images at $r^*\sim 3$~cm in Fig.\ref{VKP_picture}. Finally, the plasma potential decreases from the center to $r \sim 4$~cm (i.e. $r^*\sim 3$~cm), and presents a strong positive gradient at the edge of the plasma column. This is responsible for an $\vec{E} \times \vec{B}$ drift that drives plasma rotation in the $-\vec{e}_{\theta}$ direction, discussed later in this work.

The radial profiles of the fluctuations of the plasma parameters are shown in the middle panel of Fig.~\ref{n_Te_Vp_probe}. The fluctuations are peaked at the edge of the plasma column at $r\sim5.5$~cm (i.e. $r^* \sim 4$~cm).

Filtered light fluctuations recorded by fast camera are usually considered to be a proxy for  density fluctuations~\cite{Oldenburger_2010, Light_2013, Thakur_2014_2}. 
For the magnetic field value of $B=170$~G reported in this article, simultaneous probe and camera measurements showed the fluctuations of density $\tilde{n}$ and light intensity $\tilde{I}_{cam}$ at the probe location to have very similar spectra, as shown in the bottom panel of Fig.~\ref{n_Te_Vp_probe}.
However, as it was shown in our previous work~\cite{Vincent_2022}, for magnetic field values in the range 100 to 700~G, light intensity naturally radiated by low temperature plasmas are also highly correlated to the electron temperature. 
In Sec.~\ref{section_identification}, the comparison of experimental phase velocities with the theoretical ion acoustic speed nonetheless requires to assume $\tilde{I}_{cam}$ to be a reasonable proxy for $\tilde{n}$.
We therefore underline that this is a rather strong assumption, and that for further quantitative comparison $\tilde{I}_{cam}$ should be interpreted as a combination of both $\tilde{n}$ and $\tilde{T}_e$ - a task beyond the scope of this article.
The strong spectral component observed at $5.6$~kHz is identified as a Kelvin-Helmholtz mode~\cite{thesis_Vincent_2021}.
In the present work, we will focus on fluctuations observed between $50$ and $70$~kHz, which are unambiguously identified as ion acoustic waves.

\section{Image analysis}
\label{section_image_analysis}

The first and most natural tool that comes to mind for the images analysis is the Fourier decomposition. This is used later on; we prefer here to start the analysis by using an alternative method, the Proper Orthogonal Decomposition (POD).

Note that before performing these decomposition, we choose to normalize each pixel by its fluctuations mean, as was done in similar conditions~\cite{Thakur_2014_1}. This choice greatly enhances the contrast, allowing  to nicely extract the relative amplitudes of the modes (especially in the regions of low light intensity), but at the cost of losing information on the absolute amplitude of the modes.

Note finally that in the following the light fluctuations recorded by the camera $\tilde{I}_{cam}$ will simply be denoted $I$ for easier readability.

\subsection{Proper Orthogonal Decomposition}
\label{subsec::image_analysis_POD}

The POD consists in extracting the spatial structures that are dominant throughout time in a given data set $I(\mathbf{r^*},t)$, where $\mathbf{r^*}$ denotes space. This is done by computing the eigenmodes $\Psi_i$ of the spatial autocorrelation of the time-averaged field $\langle I \rangle(\mathbf{r^*})$. These so-called spatial modes $\Psi_i$ then define an orthonormal basis onto which the original data can be projected. This can be written:
\begin{equation*}
    I(\mathbf{r^*},t) = \sum_i \sigma_i \, a_i(t) \psi_i(\mathbf{r^*})
\end{equation*}
with $\sigma_i \, a_i(t)$ being the time evolution of the data projected on the spatial modes $\Psi_i$. Here $a_i$ and $\Psi_i$ are of norm unity; the amplitude of the various components of the decomposition are thus given by the values of $\sigma_i$.

One of the most interesting aspects of this decomposition is that it is done without any a priori on the shape of the $\Psi_i$ structures: they simply come out from the computation process, as \textit{natural} modes, contrary to Fourier analysis, which projects the data onto predefined spatial and temporal structures. Hence POD might allow the emergence of structures with physical significance that are not well described by mere Fourier modes.
Thanks moreover to its simplicity of implementation and computational speed when performed onto a discrete set of data, as is explained later, POD becomes a very attractive and efficient analysis tool for experimentalists, and has grown very popular in the last decades for the analysis of data from experiments or from numerical simulations. Note that depending on the field, this technique is also referred to as Karhunen-Loève decomposition (as a reference to the original mathematical theorem) or principal component analysis.

The set of spatial modes $(\Psi_i)$ has the property of being the optimal basis for approximating the data $I$~\cite{Berkooz_1993}: for any $N$, the norm of the projection of $I$ onto $(\Psi_i)_{1,N}$, which reads $\sum_1^N \sigma_i^2 $, is higher than the projection onto any other basis than one might choose. For a given value of $N$, the spatial modes $(\Psi_i)_{1,N}$ can then be interpreted  as the vectors that are the best suited to reproduce the information carried by I, in the most efficient way.
Applied to physical data, this property is even more interesting if the $\sigma_i^2$ have a clear physical meaning. The use of POD on experimental data has been initiated in fluid dynamics, for the analysis of turbulent velocity fields~\cite{Lumley_1967}. In this context the norm $\sum_1^N \sigma_i^2$ of the projected data represents a kinetic energy, and the modes $\Psi_i$ may then be interpreted as the most important flow structures in terms of kinetic energy.
POD has then been applied to spatio-temporal measurements of plasma fluctuations in tokamaks, either measured by sets of Langmuir probes~\cite{Benkadda_1994, VanMilligen_2015, Hansen_2015} or by means of soft x-ray emission~\cite{Dudok_1994}.
It has also been applied to camera imaging data of plasma naturally radiated light to exhibit spiral shaped structure generated by a $m=2$ instability in a linear device~\cite{Tanaka_2010} and in a tokamak to highlight plasma response to resonant magnetic perturbations~\cite{Angelini_2015}. More recently, the technique was used to characterize instabilities in the plume of a Hall thruster from fast imaging data~\cite{Desangles_2020}. Following this study, POD has been applied to decompose the camera imaging data of a plasma plume produced by a high-current hollow cathode~\cite{Becatti_2021}.
Unfortunately, in the latter cases, the physical interpretation of the $\Psi_i$ vectors is not as straightforward as in fluid mechanics, since the extracted modes result from the decomposition of light intensity fields, which depends in a non trivial way on the plasma parameters. And even by considering as a crude approximation $\tilde{I}_{cam} \propto \tilde{n}$, not much can be said on the norm $\sum_1^N \sigma_i^2$ in terms of physical significance. This does not mean the amplitude of the modes extracted from plasma emitted light is void of meaning, but simply that one has to be careful before thinking of it as a precise energy estimation. In this article POD decomposition is thus discussed in a purely qualitative way.
Finally, POD does not require any symmetry, which is a significant advantage over Fourier decomposition for instance. In the case of complex geometries, POD can be an efficient alternative for capturing the physical structures in the data. 

\begin{figure*}[t]
    \centering
    \includegraphics[width = 1.98\columnwidth, trim={0in 0in 0in 0in},clip]{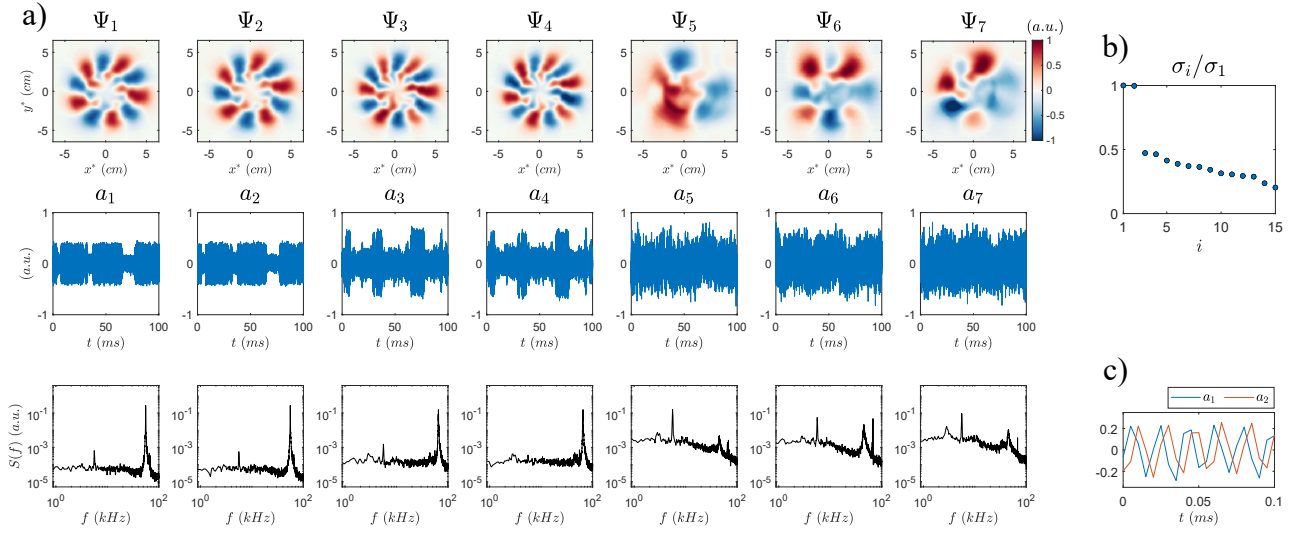}
    \caption{Proper Orthogonal Decomposition modes (a) and singular values (b) of light intensity normalized fluctuations, for $p_0=1.3$~mTorr. c) Zoom on the evolution of $a_1$ and $a_2$. See text for details.}
    \label{iaw_POD}
\end{figure*}

In practice, it can be shown that a direct extraction of the spatial modes $\Psi_i$ associated to their temporal evolution $a_i$, is in fact achieved by applying a mere singular value decomposition (SVD) to the matrix containing the data $I$, rearranged in such a way that one dimension of the matrix represents space, and the other time. This way of computing a POD, also referred to as bi-orthogonal decomposition~\cite{Aubry_1991} is the one implemented here.
Each image (of $p$ pixels) of a video containing $q$ frames is rearranged to form a matrix $I$ of size p$\times$q to which a singular value decomposition is applied $I = \Psi \Sigma A^t$. $\Psi$ and $A$ are orthonormal matrices of respective sizes $p \times p$ and $q \times q$, and $\Sigma$ is a matrix of the same size as $I$. The matrix $\Sigma$ only contains diagonal elements, which are the decomposition's singular values $\sigma_i$:
\def\tmp{
\begin{bmatrix}
    & \ & \\
     & \ & \\
     & \ & \\
     & \ (I)_{ij} \\
     & \ &  \\
     & \ & \\
     & \ & \\
\end{bmatrix}
}

\scriptsize
\begin{equation*}
\stackMath\def\stackalignment{r}%
  \stackon%
    {\mathrm{\textit{space}}\left\{\tmp\right.}%
    {\overbrace{\phantom{\smash{\tmp\mkern -35mu}}}^{\mathrm{\textstyle \textit{time}}}\mkern 20mu}%
=
\begin{bmatrix}
    \ & \ & \ & \ & \  \\
    \vdots & \vdots  & \ & \ & \vdots  \\
    \Psi_1 &  \Psi_2 & \ & \ & \Psi_q \\
    \vdots & \vdots  & \ & \ & \vdots  \\
    \ & \ & \ & \ & \  \\
\end{bmatrix}
\begin{bmatrix}
    \sigma_1 & \ &  \\
    \ & \ddots  &  \\
    \ & \ & \sigma_q  \\
    \ & \ & \  \\
    \ & \ & \  \\
        \ & \ & \  \\
\end{bmatrix}
\begin{bmatrix}
    & \cdots \ a_1 \ \cdots & \\
    & \cdots \ a_2 \ \cdots & \\
    & \ & \  \\
    & \cdots \ a_p \ \cdots & \\
\end{bmatrix} 
\end{equation*}
\normalsize

Figure~\ref{iaw_POD} a) shows the result of a POD applied to a 100~ms time series of  intensity fluctuations (i.e. 20000 images) recorded at a pressure $p_0=1.3$~mTorr. The spatial modes $\Psi_i(\mathbf{r^*})$ are displayed in the top row, the time series of the amplitudes $a_i(t)$ in the the middle row, and the corresponding power spectral density $S(f)$ in the bottom row. The interpretation of the decomposition requires to consider pairs of modes, such as $I^{POD}_{1,2}(\mathbf{r^*},t) = \sigma_1  a_1(t) \Psi_1(\mathbf{r^*}) + \sigma_2  a_2(t) \Psi_2(\mathbf{r^*}) $, which yields rotating azimuthal waves of the type $e^{-i \omega t - i m \theta }$ (shown later in subsection~\ref{subsec::comparison_POD_2DFT}), since the spatial modes $\Psi_1$ and $\Psi_2$ are shifted by a quarter wavelength and the temporal modes $a_1$ and $a_2$ are in quadrature (see Fig.~\ref{iaw_POD} c))\footnote{Due to the very high frequency of the waves (75 kHz) relative to the sampling frequency (200 kHz), the signals $a_1$ and $a_2$ are closer to triangular than sinusoidal shapes; but other measurements of lower frequency waves clearly show sinusoidal evolutions for the $a_i$ signals}. In the example shown in Fig.~\ref{iaw_POD}, the modes ($\Psi_1$, $a_1$) and ($\Psi_2$, $a_2$) correspond to a $m=-5$ rotating azimuthal wave, the modes ($\Psi_3$, $a_3$) and ($\Psi_4$, $a_4$) correspond to a $m=-6$ rotating azimuthal wave, and the modes ($\Psi_6$, $a_6$) and ($\Psi_7$, $a_7$) correspond to a $m=-4$ rotating azimuthal wave. 

The rotation frequencies of the $m$-modes can be deduced from the spectra of the temporal signals $a_i$, shown in the bottom row of Fig.~\ref{iaw_POD} a). A very clear peak at frequency $f=55.6$~kHz is observed for the modes $1 \& 2$, capturing the $m=-5$ azimuthal wave. The  $m=-6$ wave (POD modes $3 \& 4$) has a $f=65.1$~kHz frequency, and the $m=-4$ wave (POD modes $6 \& 7$) has a $f=45.2$~kHz frequency. Section \ref{section_identification} shows that these modes are ion acoustic waves.

The singular values $\sigma_i$ of the modes are plotted in Fig.~\ref{iaw_POD} b). The amplitudes of $\sigma_1$ and $\sigma_2$ are nearly identical (within $0.3\%$), and more than twice larger than the other singular values, showing that the dynamics is dominated by a $m=-5$ rotating mode.
The time series $a_i(t)$ show sudden changes in amplitude, for instance at times $t\sim3.8$~ms, $29.5$~ms and $65.2$~ms, corresponding to an energy exchange between modes $m=-6$ and $m=-5$, that will be investigated in Section \ref{section_NL_interactions}.
The relatively intense POD mode ($\Psi_5$, $a_5$), with a strong spectral component at $f=5.6$~kHz, was identified as a $m=3$ Kelvin-Helmoltz mode~\cite{thesis_Vincent_2021}, not discussed here.
The POD analysis presented here was straightforward to implement, and provide a very efficient way of extracting global features captured in a video sample.
Now we present the results of a Fourier analysis performed on the same data that complements the POD analysis.

\subsection{2D Fourier Transform}
\label{subsec::image_analysis_2DFT}

The 2D Fourier Transform (2D-FT) of a two variables function $f(\mathbf{x}, t)$ reads:  $\hat{f}(\mathbf{k}, \omega) = \iint f(\mathbf{x}, t) e^{-i(\mathbf{k}\cdot\mathbf{x}+\omega t)}d\mathbf{x}dt$.
2D-FT is classically used to decompose the spatio-temporal signals collected by azimuthally distributed probe arrays into azimuthal modes~\cite{Latten_1995, Yamada_2010}.
Following many studies using camera imaging and performed in the context of linear plasma devices~\cite{Brandt_2011, Light_2013, Brandt_2014, Ohdachi_2017} and plasma thrusters~\cite{Mazouffre_2019, Romadonov_2019}, the 2D-FT is here performed on virtual rings at various radii $r^*$. For a given value of the radius $r^*$, a time series $I(\theta, t)|_{r^*}$ is extracted from the camera images. For each angle $\theta_n = \frac{2 \pi n}{N_{\theta}}$ with $n \in [0: N_{\theta}-1 ]$, the value of the pixel at position ($r^*$, $\theta_n$) is extracted. The angle resolution of $N_\theta=700$ is chosen here, such that no interpolation is needed in the processes of converting either a ring of pixels in the image space ($x^*$, $y^*$) into a vector along the $\theta$ direction, nor in the inverse process,  when reconstructing images in the ($x^*$, $y^*$) plane from the mode decomposition results.
Since the images are $2\pi$ periodic in the $\theta$ direction, the wave-vectors read $m/r^*$, with $m$ an integer, and the 2D-FT of $I(\theta, t)|_{r^*}$ is computed as

$$ \hat{I_{r^*}}(m, f) = \iint I(\theta, t)|_{r^*}e^{-i(m\theta + 2\pi f t)}d\theta dt$$

\noindent with $f$ the frequency. The resulting 2D power spectrum $S_{r^*}(m, f) = |\hat{I_{r^*}}(m, f)|^2$ displays the amplitudes of light intensity fluctuations as a function of the spatial mode $m$ and the frequency $f$, at a given radius $r^*$.
Note that at radius $r^* \sim$ 3.5 cm on the images, the intensity $I(\theta, t)|_{r^*}$ is reconstructed from a corona of $\sim 400$ pixels, which ensures a very good precision in the extraction of the first modes $m$ up to $m \sim 20$.
The power spectrum at $r^* = 3.3$~cm and $p_0=1.3$~mTorr is shown in Fig.~\ref{iaw_DR_1,3}. The observations are similar to those drawn from the POD analysis~: the dominant mode is an $m=-5$ mode whose frequency is peaked at 55.6~kHz, and the other important modes are an $m=-6$ mode peaked at 65.2~kHz and an $m=-4$ mode peaked at 44.9~kHz. Note that this 2D power spectrum provides a dependence of the dominant frequency on the mode number, and can therefore be seen as an experimental dispersion relation. This is used in section~\ref{section_identification} for the identification of the modes.

\begin{figure}
	\centering 
    \includegraphics[width = 0.98\columnwidth, trim={0in 0in 0in 0in},clip]{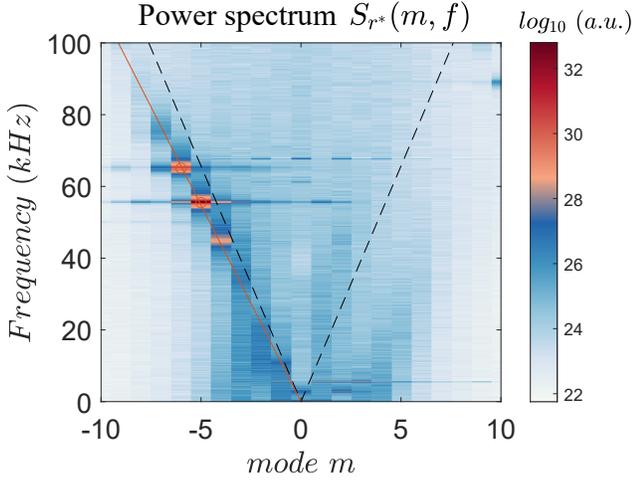}
    \caption{Power spectrum of the raw light intensity fluctuations from camera imaging, taken on a corona of radius $r^*=3.3$~cm for $p_0=1.3$~mTorr. Dispersion relations are plotted from experimental fits of the power spectrum maxima (red) and from the ion acoustic speed (black) (see section~\ref{section_identification}).}
    \label{iaw_DR_1,3}
\end{figure}

The full spatio-temporal evolution of any given $m$ mode can also be extracted by 2D-FT.
To this end, the 2D Fourier Transform is computed for radii $r^*$ covering the full image (here 2D-FT are computed for $r^* = 2i \ \mathrm{px}, \  i\in[1:64]$, instead of $r^* = [1:128]$~px, to limit memory storage and increase computational speed).
For given $m$ and $r^*$ values,  the inverse Fourier Transform of $\hat{I}_{r^*}(m, f)$ is computed, resulting in the spatio-temporal signal of the $m$ mode, at radius $r^*$. Performing this inverse computation for all radii previously mentioned leads to the full spatio-temporal reconstruction of the $m$ mode. Examples of snapshots of such reconstructed 2D-FT modes are shown and commented in subsection~\ref{subsec::comparison_POD_2DFT}. 

\begin{figure}
	\centering 
    \includegraphics[width = 0.98\columnwidth, trim={0in 0in 0in 0in},clip]{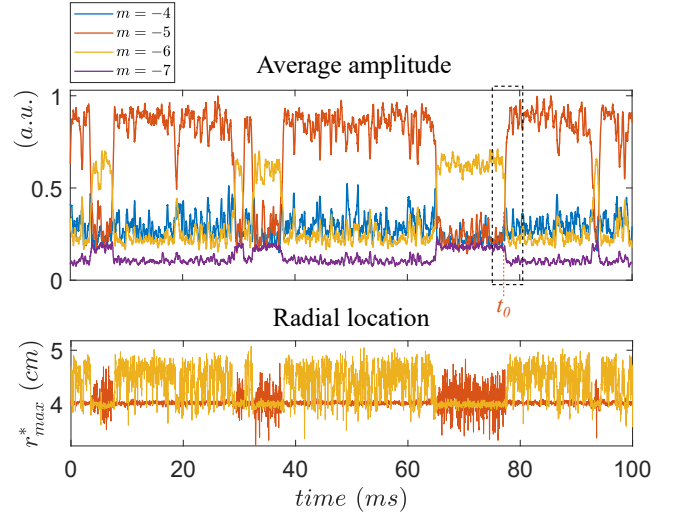}  
    \caption{Time evolution of $m$-mode average amplitudes extracted by 2D-FT, for $p_0 = 1.3$~mTorr.}
    \label{iaw_amplitude_time}
\end{figure}

The average amplitude of the reconstructed modes is then computed along time, providing a global picture of the modes dynamics.
The global $m$-modes time evolution for $p_0=1,3$~mTorr is plotted in Fig.~\ref{iaw_amplitude_time} (top).
As already observed on the time signals of the POD in Fig.~\ref{iaw_POD}, clear exchange events can be observed involving modes $m=-5$ and $m=-6$. The $m=-4$ mode is seen to follow the dynamics of $m=-5$  while the $m=-7$ mode follows the dynamics of $m=-6$ mode, a feature that was not detected by the POD analysis. From the reconstructed signals of the individual $m$-modes, the instantaneous mean radial profile is computed by an integration over $\theta$, allowing for the computation of the radial location $r^*_{max}$ where the wave amplitude is maximal. Figure~\ref{iaw_amplitude_time} (bottom) shows $r^*_{max}$ for the $m=-5$ and $m=-6$ modes, that are highly correlated to the global dynamics of the modes. Again this could not be deduced from POD, since spatial modes structure are deduced from a time-averaged analysis. Figure~\ref{iaw_amplitude_time} is further discussed in section~\ref{section_NL_interactions}. 
Now let us compare the results obtained from both POD and 2D-FT analysis.

\subsection{Comparison between POD and 2D Fourier Transform} 
\label{subsec::comparison_POD_2DFT}

\begin{figure}
	\centering 
    \includegraphics[width = 0.98\columnwidth, trim={0in 0in 0in 0in},clip]{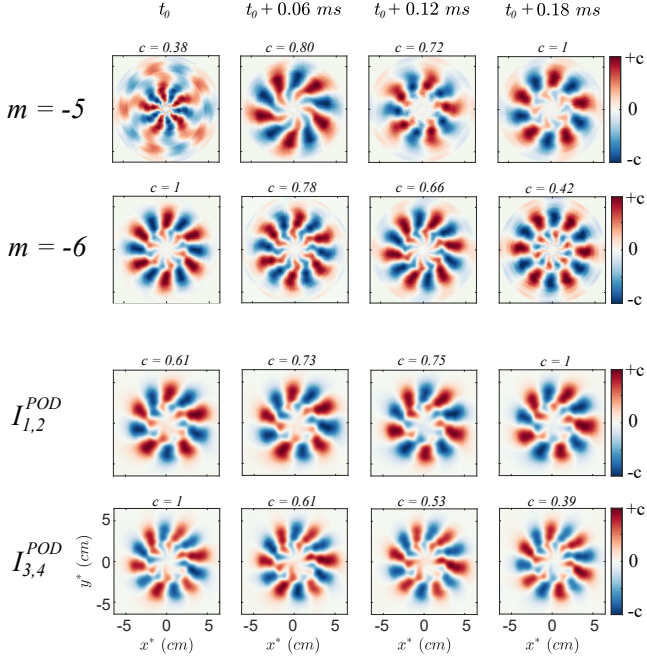}
    \caption{Evolution of 2D-FT modes $m=-5$, $m=-6$ and POD reconstructed modes $I^{POD}_{1,2}$ and $I^{POD}_{3,4}$, during the exchange event highlighted by the dashed-black box in Fig.~\ref{iaw_amplitude_time}, for $p_0=1.3$~mTorr.}
    \label{iaw_image_series}
\end{figure}

Figure~\ref{iaw_image_series} shows snapshots of the $m=-5$ and $m=-6$  2D-FT reconstructed modes, as well as the corresponding modes reconstructed from the POD analysis $I^{POD}_{1,2}$ and $I^{POD}_{3,4}$, for the experiment achieved at $p_0 = 1.3$~mTorr. 
The snapshots are shown every 0.06 ms following $t_0=77.27$~ms, marking the beginning of an energy exchange between modes $m=-5$ and $m=-6$ (see Fig.~\ref{iaw_amplitude_time} (top)). The time interval 0.06 ms represents slightly more than 3 wave periods for the $m=-5$ mode, and nearly 4 wave periods for the $m=-6$ mode. The spatial shape of the 2D-FT modes varies significantly. 
On the contrary the shapes of the POD reconstructed modes remains almost unchanged. This is actually expected since each of these signals is merely composed of the linear combination of two spatial fields. Note also that the spatial structures $\Psi_i$ were extracted from the time-averaged data field (see subsection~\ref{subsec::image_analysis_POD}): the reconstructed mode are unable to account for spatially localized variations.

\begin{figure}
	\centering 
    \includegraphics[width = 0.98\columnwidth, trim={0in 0in 0in 0in},clip]{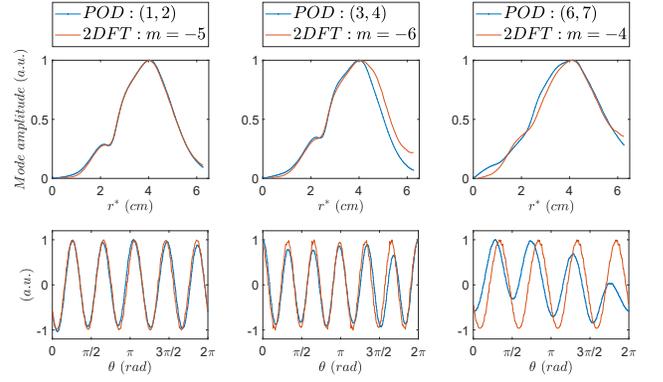}
	\caption{Comparison of (top) radial and (bottom) azimuthal profiles of the POD and 2D-FT modes, at $p_0=1.3$~mTorr, for the (left) $m=-5$, (center) $m=-6$ and (right) $m=-4$ modes.}
    \label{comparison_POD_2DFT}
\end{figure}

Let us now compare the spatial structures provided by POD and 2D-FT. Figure~\ref{comparison_POD_2DFT} (top) shows time-average radial profiles of the modes for both decompositions. The profiles are computed by an integration along $\theta$, averaged over the 20000 images. Fig.~\ref{comparison_POD_2DFT} (bottom) shows the azimuthal profiles taken at a given time and for $r^* = 4$~cm. 
The comparison between POD modes ($1 \& 2$) and the $m=-5$ 2D-FT mode  shows an almost perfect match (note that the match slightly decreases when the amplitude of the $m=-5$ mode strongly decreases). The comparison between POD modes $I^{POD}_{3,4}$ and the $m=-6$ 2D-FT mode give similar results, although with a lower agreement on the outward part ($r^* \geq 4$~cm) of the radial profiles.
An overall good match is observed between the lower amplitude POD modes ($6 \& 7$) and the $m=-4$ 2D-FT mode radial profiles. The instantaneous azimuthal profile are not identical, with a phase shift up to $\sim \pi/8$ depending on the frame.
These results show that, in the context of data having 2-$\pi$ periodicity, POD and 2D-FT decompositions share several common features, while they do provide exactly the same knowledge. 
Note finally that for the computations performed here with a number of images $N=20000$, the POD is twice faster than the 2D-FT (even though it was taken into account for the 2D-FT only 20 mode reconstructions, and half of the images pixels as mentioned in subsection~\ref{subsec::image_analysis_2DFT}). Then when only taking $N = 2000$, the POD is more than 12 times faster than the 2D-FT.

The main strengths of both POD and 2D-FT techniques are summarized:
\begin{itemize}
    \item POD is fast and easy. It is extremely simple to implement, and it provides quick and direct results on the spatio-temporal dynamics of a dataset.
    \item POD is flexible. It does not rely on any particular shape of the physical structure at play, nor on a specific location in the images analysed. It will therefore be particularly well suited to study for instance non-linearly saturated modes exhibiting a complex spatial or temporal pattern. Note however that if the results can be particularly insightful, they might also be difficult to interpret (and in some cases even unusable).
    \item 2D-FT is explicit, hence robust. Projecting the data onto a predefined set of wave modes (here for instance of the form $e^{-i \omega t -i m \theta}$) prevents the emergence of unexpected structures, but it provides the results with a well identified physical meaning.
    \item 2D-FT is exhaustive for linear mode analysis. Since it provides the full spatio-temporal evolution of linear wave, 2D-FT is particularly attractive to study their dynamics, exhibit the corresponding dispersion relations, or use for instance the phase correlations between modes to study weakly non-linear interactions (see the use of bicoherence in section~\ref{section_NL_interactions}).
\end{itemize}

Both techniques can provide insightful and complementary results.
A recent preprint, reporting on the specific comparison between POD and 2D-FT applied to Hall thruster camera imaging~\cite{Brooks_2022}, concludes similarly. Applied to the present datasets, POD shows that the dominant physical structures are $m$-modes of the form $e^{-i\omega t - i m \theta}$. This indicates that the 2D-FT as implemented here, is an appropriate numerical tool for the mode decomposition. Hence POD does not constitute a strong gain for further analysis here. In the following, for the identification of the waves and the in-depth study of their weakly non-linear interactions, we will use the results from the 2D-FT decomposition.

\section{Waves identification}
\label{section_identification}

\begin{figure}
	\centering 
    \includegraphics[width = 0.9\columnwidth]{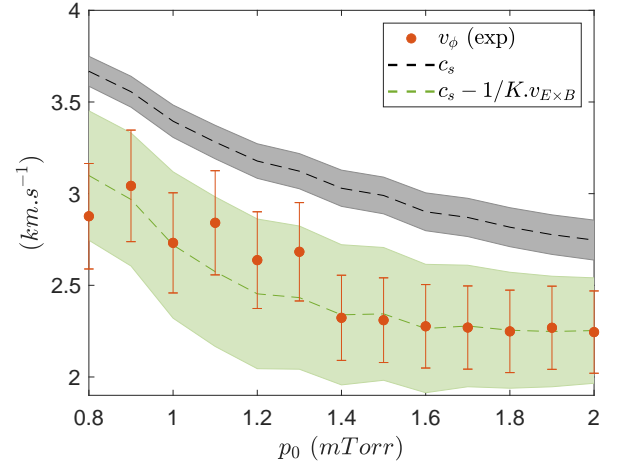}   
    \caption{Comparison between the experimental phase velocity (red dots), the ion sound velocity $c_s$ (black curve), and its Doppler shifted values (green curve).}
    \label{v_phi_IAW_exp_theo}
\end{figure}

The azimuthal waves detected by both POD and 2D-FT are now unambiguously identified as ion acoustic waves. A series of high-speed imaging acquisitions was performed for pressures $p_0$ in the range $[0.8;  2]$~mTorr by steps of 0.1 mTorr.
For each value of the pressure, the radius $r^*_{max}$ at which the wave amplitude is maximal is deduced from the time-average of the raw images.
The experimental phase velocity $v_{\phi}$ is determined by a linear fit of the most energetic modes observed on the spectrum $S_{r^*_{max}}(f,m)$ as $f(m) = v_{\phi} m / (2 \pi r^*_{max}) $. A typical linear fit is shown in Fig.~\ref{iaw_DR_1,3} for $p_0 = 1.3$~mTorr. The experimental phase velocities $v_{\phi}^{exp}$ are displayed in Fig.~\ref{v_phi_IAW_exp_theo} as red dots. The errorbars are estimated by the combination of the uncertainties on the fit on $S(m,f)$, and on the evaluation of $r^*_{max}$ ($r^*_{max}(t)$ fluctuates around its mean value with a standard deviation of $\sim 3\%$, see Fig.~\ref{iaw_amplitude_time} (bottom)).

These experimental phase velocities are compared to the theoretical ion acoustic speed $c_s = \sqrt{e T_e / m_i}$, with $e$ the elementary charge and $m_i$ the ion mass. The computation of the latter requires careful estimates of $T_e$ where the phase velocity is measured on the high-speed images.  Note that at $z=49$~cm, where the probe measurement is performed, the radial position that is best representative of what is seen at $r^*=3.3$~cm on the images is in fact at $r=5$~cm (see Appendix~\ref{appendix::radial_scale} and for a detailed explanation see \cite{Vincent_2022}). A detailed pressure scan of the electron temperature $T_e$ was performed with the 5-tips probe at a radius $r=4$~cm, and from a finely resolved radial scan at $p_0=1$~mTorr~\cite{Vincent_2022,thesis_Vincent_2021}, we have $T_e(5~$cm$) \approx T_e(4~$cm$) + 0.2$~eV. Therefore, from the measured values $T_e(4~$cm$)$, $T_e(5~$cm$)$ is evaluated to lie in the range $[T_e(4~$cm$) ;T_e(4~$cm$)+0.5]$~eV. The resulting theoretical ion acoustic speeds $c_s(p_0)$ are shown in Fig.~\ref{v_phi_IAW_exp_theo} (gray area).

The experimental phase velocities follow the trend of $c_s(p_0)$, with values shifted down by approximately $700 $~m/s. This is well explained by a Doppler shift due to the plasma column rotation. The plasma column indeed rotates, as was reported previously~\cite{Desangles_2021}, where the electric drift $\displaystyle\frac{\Vec{E}\times\Vec{B}}{B^2}= - \nabla_r \phi_p/B \, \vec{e}_{\theta}$ was shown to overcome the 
diamagnetic drift $ - \displaystyle\frac{T_i}{n} \frac{\Vec{\nabla} n \times\Vec{B}}{B^2} $. Two damping mechanisms also need to be accounted for: ion-neutral friction and effective friction due to ionization. The ion-neutral collision frequency reads $\nu_{in} = n_n \sigma_{in} v_{th,i}$, with $v_{th,i} =  \sqrt{e T_i(eV)/m_i}$, $n_n$ being the neutral density and $T_i$ the ion temperature. We consider $n_n \approx  p_0 / k_B T_n$ with $T_n(p_0)= 350$~K, $\sigma_{in} = 1.6 \times 10^{18}$~m$^{-2}$ from experimental cross sections~\cite{Phelps_1994}, and $T_i \approx 0.2$~eV  using previous LIF measurements. 
The effective friction due to the ionization originates from ions created with a temperature much lower than the surrounding $T_i$ and depends upon the ionization frequency $\nu_{iz}$, computed as $\nu_{iz} = n_n K_{iz,0} T_e^{0.59} \exp(-\varepsilon_{iz}/T_e)$, with $K_{iz,0} = 2.34 \times 10^{-14}$~m$^3$/s and $\varepsilon_{iz} = 17.44$~eV, and $T_e$ in eV~\cite{book_Lieberman}. A global damping factor $K$ is then given~\cite{Desangles_2021,thesis_Vincent_2021} as
$K = 1 + \left( \displaystyle\frac{\nu_{in} + \nu_{iz}}{\omega_{ci}} \right)^2 $, with $\omega_{ci}$ the ion cyclotron frequency.
This finally gives a background azimuthal rotation of the ions as $ v_{i0,\theta} \approx - (1/K) \partial_r \phi(r) / B $.

The rotation velocity $v_{i0,\theta}$ is estimated from the experimental profiles $\phi_p(r)$ shown in Fig.~\ref{n_Te_Vp_probe} (measured at $p_0=1$~mTorr, and assuming variations with pressure within $\pm 20\%$). The estimated values of $n_n$ and $T_i$ are considered to be bounded within $\pm 10 \%$, and  $T_e$ is estimated from $T_e^{r=4cm}(p_0)$ as explained above.
The results for the estimate of $c_s + v_{i0,\theta}$ are shown in Fig.~\ref{v_phi_IAW_exp_theo} (green curve).
In spite of all the approximations made, the comparison between experimental phase velocities and the Doppler shifted values of $c_s$ provides a very satisfactory agreement. 
This allows us to identify with great confidence the azimuthal waves observed at $B=170$~G as ion acoustic waves.
An interesting feature is that the ion acoustic waves travel in the positive $\theta$ direction, i.e. opposite to the $E \times B$ drive.

We stress here that adding the ion background velocity to the classical ion acoustic wave speed is a crude approximation, deemed sufficient here for the purpose of wave identification. However, a careful calculation would require to compute a complete dispersion relation from the governing equations, which couple in a complex way and prescribe direct analytical computation. Indeed the effect of an ion background velocity on the ion acoustic phase velocity is likely to be coupled with other effects such as electron magnetization or friction with the neutrals, leading to computations well beyond the scope of this article.

Interestingly, we observed that the ion acoustic waves are only observed over a narrow range of magnetic field values. For $B=80$~G no clear wave emerges from the fluctuations of the plasma density or emitted light intensity; on the other hand for $B \geq 300$~G low frequency waves develop~\cite{thesis_Vincent_2021}.

\section{Modes dynamics and interactions}
\label{section_NL_interactions}

The spatio-temporal dynamics and the non-linear nature of the energy exchanges between the ion acoustic modes, as clearly shown in Fig.~\ref{iaw_amplitude_time}, are now described.

\subsection{Growth rates of ion acoustic modes}
\label{subsec::exchange_dynamics}

The time series shown in Fig.~\ref{iaw_image_series} is taken around the exchange event highlighted at $t_0$ in Fig.~\ref{iaw_amplitude_time}.
At time $t_0$ the amplitude of the $m=-6$ mode is close to its maximum, while the amplitude of the $m=-5$ mode is close to its minimum. At time $t_0 + 18$~ms, the  amplitude of the $m=-6$ mode has decreased close to its minimum value, and the $m=-5$ mode dominates.
Figure~\ref{iaw_amplitude_time} (bottom) shows that the radial position of the dominant mode (either the $m=-5$ or the $m=-6$ mode) is indeed very stable. On the other hand, the radial position of the low amplitude mode strongly fluctuates around its equilibrium value (with standard deviations around $~0.6$~cm for the $m=-5$ mode and $\sim 0.5$~cm for the $m=-6$ mode).

\begin{figure}
	\centering 
    \includegraphics[width = 0.95\columnwidth]{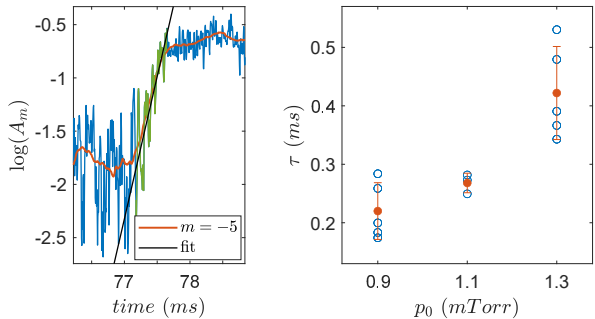}
    \caption{Left: time evolution of $m$-modes average amplitudes, extracted by 2D-FT. Left: fit of the 2D-FT $m=-5$ mode growth rate at an exchange event with $m=-6$ mode, for $p_0 = 1.3$~mTorr. The fit is done on a selected interval of the raw data (light blue). The red curve is the result of a filter. Right: Evolution of the growth time scale of $m=-5$ mode, evaluated during exchange events, as a function of the pressure $p_0$.}
    \label{iaw_m_mode_exchanges_time}
\end{figure}

The exchange events observed for $p_0=1.3$~mTorr between modes $m=-5$ and $m=-6$ (Figures~\ref{iaw_POD} and~\ref{iaw_amplitude_time}) are similarly observed at $p_0=1.1$~mTorr and  $p_0=0.9$~mTorr. The timescales of the exchange events are now determined at these three values of the pressure. This is done by fitting the mode amplitude $A_m$ as  exponentially growing: $A_m \propto \exp(t/\tau)$. Figure~\ref{iaw_m_mode_exchanges_time} (left) shows a typical  fit around $t_0$: the green part shows the interval over which the raw signal (blue) is fitted; a low-pass filtered signal is shown for clarity (red). Figure~\ref{iaw_m_mode_exchanges_time} (right) shows the resulting values of $\tau$ found for the $m=-5$ mode. The growth-time $\tau$ significantly increases with the pressure, its value doubles from $p_0=0.9$~mTorr to $1.3$~mTorr. This is interpreted as being the result of an increased friction from the neutrals at higher pressure. Note that this observation of a decrease of the ion acoustic wave growth rate with increasing pressure is consistent with theoretical predictions~\cite{Baalrud_2016}.

\subsection{Residence time distribution}
\label{subsec::exchange_dynamics}

\begin{figure}
	\centering 
    \includegraphics[width = 0.8\columnwidth, trim={0in 0in 0in 0in},clip]{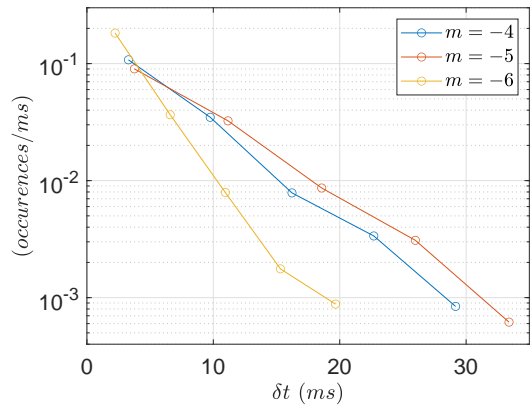}
    \caption{Residence time PDF, obtained for a neutral pressure $p_0 = 1.30$~mTorr.}
    \label{bistability}
\end{figure}

The statistics of the transitions between the $m=-5$ and $m=-6$ modes were obtained in a new set of experiments, performed at a lower sampling frequency (20 kfps)~\footnote{Note that with the lower sampling frequency of 20 kfps, the extracted IAW modes with frequencies $\sim 70$ kHz can not be resolved temporally, which prevent the distinction between modes $+m$ and $-m$ for a given integer $m$. However it is observed that in the same conditions, with higher sampling frequency acquisition, the (m=+p) mode amplitude is negligible in front of the (m=-p) amplitude for $p=[4, 7]$. We therefore use at $F_s = 20$~kfps the sum of the amplitudes of extracted modes +p and -p, to estimate the amplitude of mode (m = -p),  for $p=[4, 7]$.} over longer times (2 seconds).
This allows to extract the time evolution of the modes average amplitude, extracted by 2D-FT. The probability distribution function of the residence time of the $m=-4$, $m=-5$ and $m=-6$ modes are shown in Fig.~\ref{bistability}, for a total duration of 4 seconds (i.e. more than one thousand transitions between modes).  The distributions are compatible with an exponential distribution, which implies that the transition events are not correlated. Such distributions of residence times or waiting times are ubiquitous to transitions observed in aerodynamics~\cite{Gayout2021}, turbulent flows~\cite{Ravelet2004,DeLaTorre2007} or convection~\cite{Brown2006}, to the waiting time between reversals in dynamo experiments~\cite{Monchaux2009}, or the turbulent dynamics of the scrape-off layer in tokamaks~\cite{Garcia2015,Theodorsen2018}.

For all modes, the probability distribution function is compatible with a functional fit of the form $e^{-t/\tau}$, with $\tau = 5.4$~ms for $m=-4$, and $\tau = 6.0$~ms for $m=-5$ and $\tau = 3.2$~ms for $m=-6$. As already observed in Fig.~\ref{iaw_amplitude_time}, the $m=-4$ mode is tied to the $m=-5$ mode, resulting in similar pdf. Figure~\ref{iaw_amplitude_time} also shows that the system is more often dominated by a $m=-5$ mode, which results in an exponential pdf with a larger characteristic time for the $m=-5$ mode as compared to the $m=-6$ mode.   High speed imaging of the dynamics of the plasma allows to probe long-time statistics of the waves dynamics. It opens the possibility to probe the evolution of the characteristic residence time as a function of the control parameters (for instance pressure), possibly shedding light to the physical processes leading to exchange events. Note that the dominant mode (and the associated characteristic time) was observed to strongly evolve with pressure (data not shown and beyond the scope of this article).

\begin{figure}
	\centering
    \includegraphics[width = 0.9\columnwidth, trim={0in 0in 0in 0in},clip]{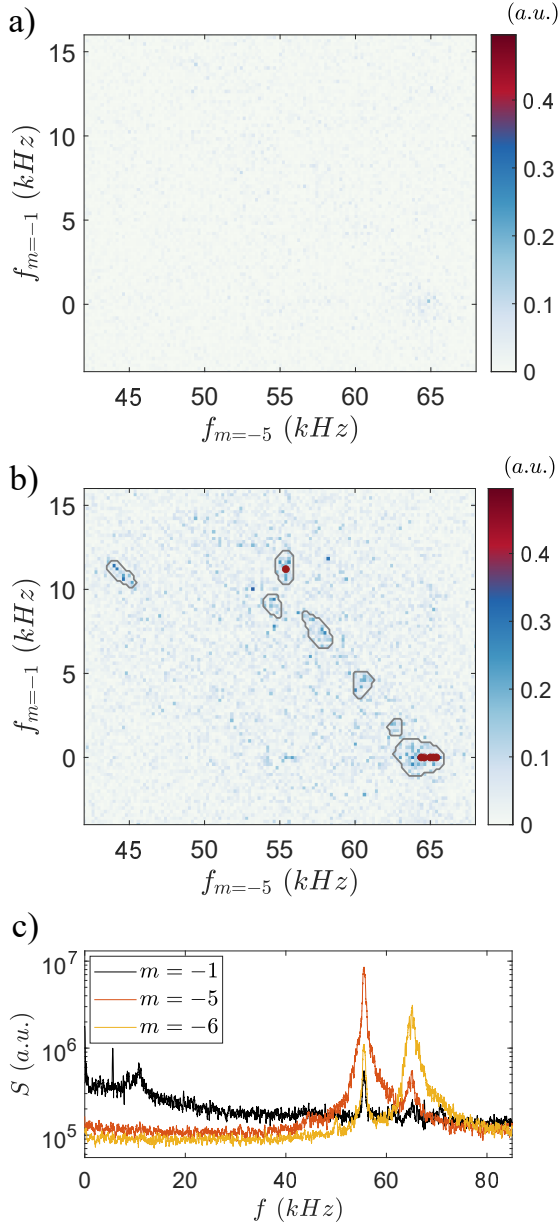}
    \caption{Maps of the threshold $b_0^2$ (a) and bicoherence $b^2$ (b), corresponding to the three-wave interaction $(m=-5) + (m=-1) \leftrightarrow (m=-6)$. c) Frequency power spectra of modes $m=-1$, $m=-5$ and $m=-6$ from 2D-FT computed at $r^* = 3.3$~cm. $B=170$~G and $p_0=1.3$~mTorr.}
    \label{bico_IAW}
\end{figure}

\subsection{Non-linear behaviour}
\label{subsec::nl_behaviour}

In order to further assess the  non-linear
nature of the dynamics between the dominant ion acoustic modes, the bicoherence $b^2(f_{m=-5},f_{m=-1})$ is computed for the three wave interaction 
$(m=-5) + (m=-1) \leftrightarrow (m=-6)$, and shown in Fig~\ref{bico_IAW}. Note that to increase statistics, the 2D-FT at all radii $1 \leq r^* \leq 5$ around the wave maximal amplitude are used. More details on the bicoherence computations are provided in  appendix~\ref{appendix::bico}. The threshold map shown in Fig~\ref{bico_IAW} a) was computed using a basic surrogate technique where the phases of the 2D-FT signal are randomly mixed. This yields bicoherence values for signals without any preferential phase relations, from which a threshold value of $\max(b_0^2) = 0.12$ is estimated. Figure~\ref{bico_IAW} b) shows the map of $b^2(f_{m=-5},f_{m=-1})$ with $f_{m=-5}$ and $f_{m=-1}$ the frequencies of modes $m=-5$ and $m=-1$ respectively. For the sake of visibility, the areas of bicoherence high values are highlighted by gray contours (defined at $40 \%$ of the maximum value of a Gaussian filtered $b^2$ map). Most of the bicoherence highest values lie around the diagonal $f_{m=-5} + f_{m=-1} = 65$~kHz, that is the dominant frequency of the $m=-6$ mode. This reveals the strong non-linear behaviour of the ($m=-6$, $f\sim 65$~kHz) mode component, which interacts with $m=-5$ and $m=-1$ modes via continuous sets of frequencies. 
The points displayed as red dots in Fig.~\ref{bico_IAW} b) are also enlarged for clarity: they correspond to $b^2 \gtrsim 0.36$, i.e. more than three times the threshold value.
The points for which $f_{m=-1} = 0$~kHz, and $f_{m=-5} \in [64.6;65.4]$~kHz correspond to frequency components of the $m=-5$ mode being fed by the high amplitude of the ($m=-6$, $f\sim 65$~kHz) component.
Note that these interactions are not the dominant process characterizing the energy exchanges detailed in subsection~\ref{subsec::exchange_dynamics}, since they only involves frequency components of the $m=-5$ mode around 65 kHz, with a low energy.
The point at $f_{m=-5} = 55.4$~kHz and $f_{m=-1} = 11.2$~kHz however corresponds to the interaction:
\begin{equation*}
\mbox{\footnotesize $(m=-5,f=55.4) \ + \ (m=-1,f=11.2) \  \leftrightarrow \ (m=-6,f=66.7)$
}
\end{equation*}
which involves the dominant frequency components of the $m=-5$ and $m=-6$ modes. The very high bicoherence value at this location ($b^2 = 0.39$) definitively establishes the non-linearity of the interactions between the ion acoustic modes  $(m=-5,f=55.4\mathrm{~kHz})$ and $(m=-6,f=66.7\ \mathrm{kHz})$, at the origin of the transitions observed in Fig.~\ref{iaw_amplitude_time}.

Finally, Fig.~\ref{bico_IAW} c) shows the frequency spectra of the $m=-6$, $m=-5$ and $m=-1$ modes involved in the three-wave interactions described above. These spectra correspond to 1D cuts along the frequency axis of the 2D-FT spectrum shown in Fig.~\ref{iaw_DR_1,3}. These spectra clearly display the non-linear feeding of the $m=-5$ mode by the high amplitude $m=-6$ mode around 65 kHz. The non-linear feeding of modes $m=-1$ and $m=-6$ by the high amplitude $m=-5$ mode around 55 kHz is also visible. The component $(m=-6, f=55\mathrm{~kHz})$ is then non-linearly interacting with $(m=-5, f=44)\mathrm{~kHz}$ and $(m=-1, f=11\mathrm{~kHz})$, as can be deduced by the high values of $b^2(f_{m=-5} \sim 44, f_{m=-1} \sim 11)$ from Fig.~\ref{bico_IAW} b).

The computation of other bicoherence maps (not shown here) reveals additional non-linear behaviours. The bicoherence computation of the  $(m=-6) + (m=-1) \leftrightarrow (m=-7)$ coupling unambiguously shows that $(m=-6,f=65.2\ \mathrm{kHz})$ non-linearly interacts with $(m=-7,f=74.0\ \mathrm{kHz})$ via an $(m=-1,f=8.8\ \mathrm{kHz})$ mode component (with $b^2 = 0.36 > 3\max(b_0^2)$). Similarly,  bicoherence computation of the $(m=-4) + (m=-1) \leftrightarrow (m=-5)$ coupling highlights that $(m=-4,f=44.2\ \mathrm{kHz})$ and $(m=-5,f=55.4\ \mathrm{kHz})$ modes non-linearly interact via $(m=-1,f=11.2\ \mathrm{kHz})$ (with $b^2 = 0.38 > 3\max(b_0^2)$). As a last example, the bicoherence map for the interaction $(m=-4) + (m=-2)  \leftrightarrow (m=-6)$ does not exhibit high values indicating the absence of non-linear interaction between the corresponding ion acoustic modes. It however reveals that the frequency components $(f=55\ \mathrm{kHz})$ of $(m=-4)$ and $(m=-6)$  modes (resulting from the spread of the $m=-5$ mode, visible in Fig.~\ref{iaw_DR_1,3}) are non-linearly linked via the $(m=-2,f=0\ \mathrm{kHz})$ mode component.

Thanks to the rich spatio-temporal information provided by camera imaging and to the use of bicoherence, the weakly non-linear interactions are clearly highlighted. In particular the existence of three-wave interactions between ion acoustic modes $m=-p$,  $m=-p-1$ and  $m=-1$, for $p\in[4,5, 6]$, is demonstrated.

\section{Conclusion}\label{sec:ccl}

We have presented the first report of temporally and spatially entirely resolved ion acoustic waves in a magnetized plasma column. The ion acoustic waves were observed by means of fast camera imaging in a low temperature argon plasma column, with dominant azimuthal mode numbers $m=-4$, $m=-5$ and $m=-6$ depending on the neutral pressure that was varied from 0.8~mTorr to 2~mTorr.

Two image analysis techniques, namely  proper orthogonal decomposition (POD) and 2D Fourier transform (2D-FT), were presented and thoroughly compared. These tools are found to be complementary. POD is easy to implement and adaptable to any type of data, and useful to provide a fast overview of the underlying dynamics of a given dataset. This helps focusing in a second time on a more precise and targeted analysis, that 2D-FT can then provide, yielding detailed and unambiguous information.

Using 2D-FT analysis of high speed images, the ion acoustic waves were found to rotate in opposite direction to the global $E \times B$ drift of the plasma column, with a phase velocity Doppler shifted by this actual electric drift velocity.

The dynamics of the dominant ion acoustic modes was then explored using the 2D-FT decomposition. Growth rates, which extraction was made possible by the camera high temporal resolution, were found to decrease as pressure increases, following previous numerical predictions.
A detailed analysis was then carried out in the particular case of $p_0 = 1.3$~mTorr. At this pressure the exchange dynamics between dominant modes $m=-5$ and $m=-6$ was shown to be of a bistable nature.
More generally the weakly non-linear nature of the $m=-p$ and $m=-p-1$ mode interaction ($p\in[4,5, 6]$), involved in a three-wave interaction with a $m=-1$ mode, was demonstrated by means of bicoherence computation.

Finally we emphasize that, except from probe measurements that were needed for the wave identification, all the results that were presented exclusively rely on fast camera imaging measurements.
This work can therefore be considered as a case study demonstrating the very powerful capabilities of fast camera imaging as a plasma diagnostics, notably for the exploration of complex waves dynamics.

\section*{Acknowledgements}

This work was partly supported by the French National
Research Agency under Contract No. ANR-13-JS04–0003-01. We
acknowledge support from the CNRS for the acquisition of the
high-speed camera and useful discussions with V. Désangles and G.
Bousselin and warmly thank P. Borgnat for advises on surrogate techniques.

\section*{Author Declarations}
\subsection*{Conflict of Interest}
The authors have no conflicts to disclose.

\section*{Data availability}
The data that support the findings of this study are available
from the corresponding author upon reasonable request.

\appendix

\section{Radial scale: camera imaging v.s. probe}
\label{appendix::radial_scale}

The magnetic field ripple and parallax in our experimental set-up leads the camera lines of sight to cross regions of different plasma parameters. The light recorded by camera, resulting of an integration process along these lines of sight, cannot be directly compared to probe measurements that are performed at $z = L_2$.

A transformation is implemented, modeling the integration along the camera lines of sight of any plasma parameter that is measured at $z = L_2$. The details of this transformation are provided in Ref.~\cite{Vincent_2022}.

\begin{figure}[h!]
	\centering 
    \includegraphics[width = 0.9\columnwidth]{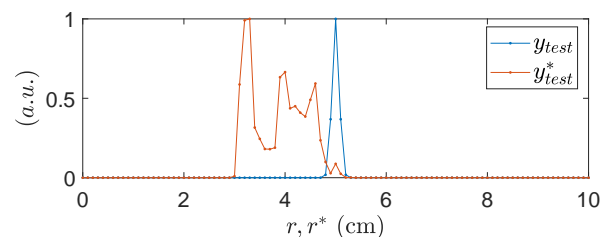}   
    \caption{Camera lines of sight integration process, applied to a test profile measured at $z = L_2$.}
    \label{y_test}
\end{figure}

Figure~\ref{y_test} shows the result of this artificial integration process, applied to a test profile peaked at $r=5$~cm (blue curve). The resulting profile (red curve), expressed along the camera imaging coordinate $r^*$, shows that what is seen on the camera images at $r^* = 3.3$~cm mainly corresponds to the plasma parameter evolution that is located at $r = 5$~cm on the axis $z = L_2$.

\section{Bicoherence and confidence level}
\label{appendix::bico}

Bicoherence is a spectral analysis tool that is commonly used in physics, for the detection of non-linear three waves interactions. Bicoherence computation essentially consists in extracting the frequency components of one of several signals, and comparing their phases. The signal decomposition at the basis of a bicoherence analysis can be done by Fourier transform~\cite{Ritz_1988, Itoh_2017} as it is the case in this work, or based on a wavelet approach~\cite{VanMilligen_1995, Oldenburger_2011}. In this appendix, we first remind the basic principle of bicoherence, and then explained how bicoherence is computed in the particular case of camera images. Then we provide the definition of a clear and mathematically meaningful threshold, that is often lacking when bicoherence is used onto experimental data in plasma physics.

\subsubsection*{Evaluate three wave interactions by bicoherence}

Let us consider three signals $x$, $y$ and $z$ with corresponding Fourier transforms $\hat{x}$, $\hat{y}$ and $\hat{z}$. 
The cross bispectrum of $x$, $y$ and $z$ is defined as a function of frequencies $(f_1, f_2)$ as~: $B(f_1, f_2) = \hat{x}(f_1) \hat{y}(f_2) \hat{z}^*(f_1 + f_2)$. 
If the frequency components $f_1$ and $f_2$ of $x$ and $y$ respectively (with phases $\phi^x_1$ and $\phi^y_2$) are involved in a three-wave interaction with the frequency component $f_1 + f_2$ of $z$ (with a phase $\phi^z_{1+2}$) the phase difference between these signals is a constant. Computing the bispectrum onto successive reduced parts $\delta_t$ of the signals is therefore a way of measuring this phase locking, since $B_{\delta t}(f_1, f_2) \propto \exp^{-i(\phi^x_1 + \phi^y_2 - \phi^z_{1+2})}$. The bicoherence is defined by the normalized average over a statistically significant number of such bispectrum computations: 
\begin{equation*}
b^2(f_1,f_2) = \frac { | \langle \hat{x}(f_1).\hat{y}(f_2).\hat{z}^*(f_1 + f_2) \rangle_{ \delta_t}|^2}{ \langle |\hat{x}(f_1).\hat{y}(f_2)|^2 \rangle_{\delta_t} \langle |\hat{z}(f_1+f_2)|^2 \rangle_{\delta_t}}
\end{equation*}
If the signal frequency components previously mentionned are perfectly uncorrelated, $b^2$ corresponds to the average of random complex numbers, and tends to cancel out. If those frequency components are on the contrary perfectly phase locked, the computations of $B_{\delta t}(f_1, f_2)$ have a constant value and $b^2 = 1$. In the case of experimental data, neither case is realistic, and a threshold value $b_0^2$ above which the bicoherence can be considered significant needs to be defined (see last paragraph of this appendix).

\subsubsection*{Bicoherence on camera images}

With camera images that provide 2D spatio-temporal signals, bicoherence can be computed between the frequency components of distinct modes $m$. Bicoherence allows to probe the  phases of  signal components for given set of wave vector and frequency ($m,f$). This analysis is applied on the present camera images, following the work of Ref.~\cite{Yamada_2010}. For a given radius $r^*$, let us denote the 2D Fourier decomposition of the light intensity:
$$ A(t,\theta) =  \sum_{n,p} a(f_n, m_p) e^{i (2 \pi f_n t - m_p \theta + \phi_{n,p})} $$

The spectrum associated with a single $m_p$ mode is a part of this decomposition:
$$ \hat{A}_{m_p}(f_n) = a(f_n, m_p) e^{i \phi_{n,p}} $$

Similarly to computations achieved for 1D signals, the bispectrum is defined as a statistical averaging, over parts of lengths $\delta t$ of the signal. In order to improve the statistical averaging here, the sum is also done over the signals from various radii $r^*$. This double averaging process is denoted $\langle . \rangle_{r^*, \delta_t}$. The bispectrum between components ($m_1$,$f_1$) and ($m_2$,$f_2$) is then defined as $B_{m_1, m_2}(f_1,f_2) = \hat{A}_{m_1}(f_1).\hat{A}_{m_2}(f_2).\hat{A}_{m_1+m_2}(f_1 + f_2)^*$, and the bicoherence is computed as:

\footnotesize
\begin{equation*}
b_{m_1, m_2}^2(f_1,f_2) = \frac { | \langle \hat{A}_{m_1}(f_1).\hat{A}_{m_2}(f_2).\hat{A}_{m_1+m_2}^*(f_1 + f_2) \rangle_{r^*, \delta_t}|^2 }{ \langle |\hat{A}_{m_1}(f_1).\hat{A}_{m_2}(f_2)|^2 \rangle_{r^*, \delta_t} \langle |\hat{A}_{m_1 + m_2}(f_1+f_2)|^2 \rangle_{r^*, \delta_t} }
\end{equation*}
\normalsize

The bicoherence as it is implemented in our code takes mode numbers $m_1$ and $m_2$ as an entry and explores all possible three-wave interactions $(m_1, f_1) + (m_2,f_2) \leftrightarrow (m_1 + m_2, f_1 + f_2)$ in terms of frequencies $f_1$ and $f_2$. The operation is fixed as an addition, and the result is in a form of a 2D map of $b_{m_1, m_2}^2(f_1, f_2)$, with $[f_1, f_2] \in [0, F_s/2]^2$, $F_s$ being the data sampling frequency. Here for simplicity, the bicoherence applied to camera images is simply denoted $b^2$. 

\subsubsection*{Definition of a threshold}

The phase correlation between any set of experimental signals is likely to be imperfect or partial, leading to $0<b^2<1$. Moreover the absolute values of the bicoherence are relative to each set of signals investigated: a general threshold value is not relevant. A method to systematically determine the level above which the value of $b^2$ becomes physically meaningful, that depends on each bicoherence computation, is therefore needed.

A possible method consists in the creation of an artificial set of signals, sharing the same characteristics than the original signals, but without any preferential relation between its frequency components. The bicoherence of this artificial set of signals is then computed, providing a lower limit for the values of $b^2$. This type of method is called surrogate technique~\cite{Siu_2009}, and can be very sophisticated. Here we use a very basic version of the surrogate techniques: the phases of each 2D-FT spectra are randomly mixed. The bicoherence computation applied to this modified data defines a threshold map $b_0^2(f_1,f_2)$. Then for simplicity we take the maximal value $\max(b_0^2)$ and define it as a global threshold value for the real bicoherence computation $b^2(f_1,f_2)$ of this dataset.

\bibliography{mybib_iaw}{}
\bibliographystyle{unsrt}

\end{document}